\documentclass[prl,aps,twocolumn,showpacs,amsmath,amssymb]{revtex4}

\usepackage{graphicx}
\usepackage{dcolumn}
\usepackage{bm}

\begin{document}
\title{Magnetic phase diagram in Eu$_{1-x}$La$_x$Fe$_2$As$_2$ single crystals}
\author{T. Wu, G. Wu, H. Chen, Y. L. Xie, R. H. Liu, X. F. Wang}
\author{X. H. Chen}
\altaffiliation{Corresponding author} \email{chenxh@ustc.edu.cn}
\affiliation{Hefei National Laboratory for Physical Science at
Microscale and Department of Physics, University of Science and
Technology of China, Hefei, Anhui 230026, P. R. China\\ }
\date{\today}

\begin{abstract}
{We have systematically measured resistivity, susceptibility and
specific heat under different magnetic fields (H) in
Eu$_{1-x}$La$_x$Fe$_2$As$_2$ single crystals. It is found that a
metamagnetic transition from A-type antiferromagnetism to
ferromagnetism occurs at a critical field for magnetic sublattice of
$Eu^{2+}$. The jump of specific heat is suppressed and shifts to low
temperature with increasing H up to the critical value, then shifts
to high temperature with further increasing H. Such behavior
supports the metamagnetic transition. Detailed H-T phase diagrams
for x=0 and 0.15 crystals are given, and possible magnetic structure
is proposed. Magnetoresistance measurements indicate that there
exists a strong coupling between local moment of $Eu^{2+}$ and
charge in Fe-As layer. These results are very significant to
understand the underlying physics of FeAs superconductors.}
\end{abstract}


\maketitle
\newpage

The discovery of superconductivity\cite{yoichi,chenxh,chen,ren} in
iron-pnicitides $LnFeAsO_{1-x}F_x$ (Ln=La, Sm, Ce and Pr) has
generated much interest for extensive study on such iron-based
superconductors, which is the second family of high-$T_c$
superconductors except for the high-$T_c$ cuprates. The magnetic
ordering of the rare earth ions $Ln^{3+}$ at low temperature has
been observed by neutron scattering\cite{rare1, rare2, rare3} except
for the spin density wave arose from $Fe^{2+}$. The coupling between
$Ln^{3+}$ and Fe$^{2+}$ has been found above ordering temperature
for local moment of rare earth ions $Ln^{3+}$\cite{rare1}. These
results indicate that the coupling between spins of rare earth ions
and Fe$^{2+}$ ions seems to be one important ingredient to
understand magnetic properties at low temperatures. It is well known
that spin density wave (SDW) is suppressed, while superconductivity
emerges with doping\cite{dong, liu, chenhong, Luetkens, cruz}.
However, no evidence is given for how to couple between SDW and
magnetic ordering of rare earth ions $Ln^{3+}$, and effect of
magnetic ordering of $Ln^{3+}$ on superconductivity. Therefore, the
coupling between the SDW from $Fe^{2+}$ and magnetic ordering of
rare earth ions $Ln^{3+}$ should be a very interesting issue. It
maybe shed light to understand the underlying physics in Fe-As
compounds.

EuFe$_2$As$_2$ is one of parent compounds with $ThCr_2Si_2$-type
structure. It shows a SDW transition around 190 K, and an
antiferromagnetic transition of Eu$^{2+}$ ions occurs at
T$_N$$\sim$20 K\cite{ren1}. Superconductivity at $\sim$ 30 K can be
achieved by K or Na doping\cite{Jeevan, Qi}. This compound is
believed to be more complicated than other parent compound due to
the large local moment of Eu$^{2+}$ ions. Similar to electron-type
$Nd_{2-x}Ce_xCu_2O_{4-\delta}$\cite{NCO1, NCO2, NCO3, NCCO1, NCCO2,
tao}, the interaction between magnetic moments of Fe$^{2+}$ and
Eu$^{2+}$ may lead to rich physical phenomena, and these maybe shed
light to the mechanism of superconductivity in these materials. In
this paper, we have studied magnetic transition by resistivity,
susceptibility and specific heat in Eu$_{1-x}$La$_x$Fe$_2$As$_2$
single crystals. The magnetic structure of Eu$^{2+}$ ions is found
to be strongly dependent on external magnetic field. A metamagnetic
transition from A-type antiferromagnetism to ferromagnetism is
observed at a certain magnetic field. The results show that the
critical magnetic field is anisotropic. With La doping, the SDW is
strongly suppressed and the critical magnetic field induced
metamagnetic transition decreases. A detailed H-T phase diagram is
given, and possible magnetic structure of Eu$^{2+}$ ions is
proposed. It is found that there exists strong coupling between
antiferromagnetic SDW in Fe-As layer and magnetic ordering of
$Eu^{2+}$, the internal magnetic field from ferromagnetic ordering
of $Eu^{2+}$ ions can polarize the antiferromagnetic SDW in Fe-As
layer.

High quality single crystals with nominal composition
Eu$_{1-x}$La$_x$Fe$_2$As$_2$ (x=0, 0.4 and 0.5) were grown by
self-flux method as described for growth of BaFe$_2$As$_2$ single
crystals with FeAs as flux\cite{Wang}. Many shinning plate-like
Eu$_{1-x}$La$_x$Fe$_2$As$_2$ crystals were obtained. The typical
dimensional is about 4 x 4 x 0.05 mm$^{3}$. Elemental analysis of
the samples was performed using energy dispersive x-ray spectroscopy
(EDX). The obtained actual La content is 0.15 and 0.18 for the
samples with x=0.4 and 0.5, respectively. The c-axis parameter is
determined by single crystal x-ray diffraction pattern (XRD). The
XRD results show that c-axis parameter shrinks with La doping from
12.13 $\AA$ for x=0 to 12.03 $\AA$ x=0.18.

\begin{figure}[t]
\includegraphics[width=9cm]{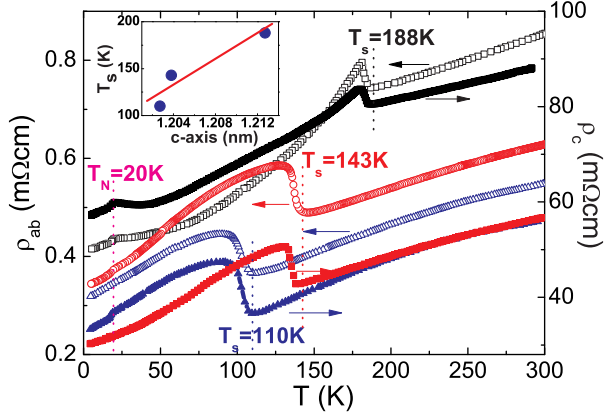}
\caption{(color online). Temperature dependence of in-plane and
out-of-plane resistivity for Eu$_{1-x}$La$_x$Fe$_2$As$_2$ single
crystals with x=0 (squares), 0.15 (circles) and  0.18
(up-triangles). The
inset shows c-axis parameter dependent SDW/structural transition temperature.\\
}
\end{figure}

Temperature dependence of in-plane and out-of-plane resistivity for
x=0, 0.15 and 0.18 crystals is shown in Fig.1. Both in-plane and
out-of-plane resistivity show similar temperature dependent
behavior. In-plane and out-of-plane resistivities for parent
compound show almost a linear temperature dependence above $\sim
188$ K, and a steep increase at 188 K, then changes to metallic
behavior. This transition is ascribed to SDW/structural
transition\cite{Wang}. With La doping, the SDW/structural transition
is suppressed with decreasing T$_s$ from 188 K to 110 K for the
crystal with x=0.18. It suggests that electrons are introduced into
the system with La doping, and lead to a decrease of $T_s$. For all
samples, there exists a kink in resistivity around 20 K. Such kink
is ascribed to antiferromagnetic transition of Eu$^{2+}$ ions. It
suggests that there exists a coupling between the local moment of
Eu$^{2+}$ ions and conducting electron in Fe-As layer.

\begin{figure}[b]
\includegraphics[width=9cm]{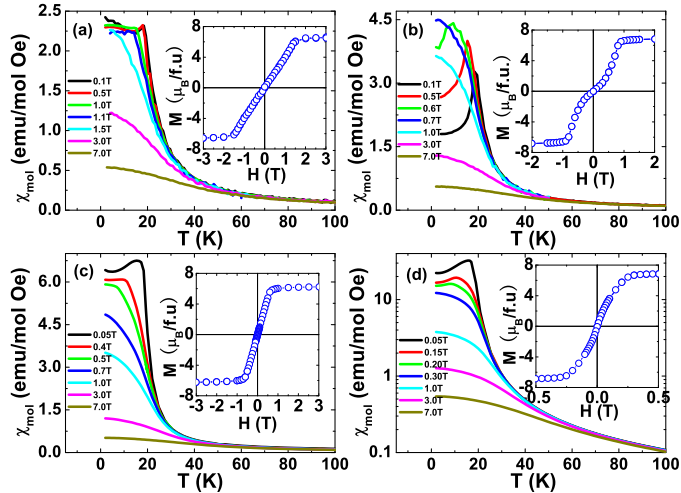}
\caption{(color online). Temperature dependence of susceptibility
measured in field-cooled process under different H, (a):
H$\parallel$c and (b): H$\perp$c for x=0 crystal; (c): H$\parallel$c
and (d): H$\perp$c for x=0.15 crystal. The inset shows M-H curves
for x=0 and x=0.15 at 2 K.
\\
}
\end{figure}

Temperature dependence of susceptibility ($\chi$) for the crystals
with x=0 and 0.15 measured in field cooled process under different H
up to 7 Tesla applied within ab-plane and along c-axis is shown in
Fig.2, respectively. For the crystal with x=0, a standard
Curie-Weiss behavior is observed in high temperature region (T $>$
50 K) for both H$\parallel$ab and H$\perp$ab. Below 20 K, a steep
decrease in $\chi$ happens with H$\parallel$ab plane, while the
$\chi$ with H$\perp$ab plane emerges a small peak, then almost
remains constant. It suggests that an antiferromagnetic transition
occurs below 20 K. For x=0.15 crystal, temperature dependent
susceptibility at high temperature is also Curie-Weiss behavior. The
antiferromagnetic transition shows up at T$_N$ of 16 K. All these
phenomena occur only at low magnetic fields. With increasing H
larger than a critical magnetic field, the antiferromagnetic
character in $\chi(T)$ disappears as shown in Fig.2. As shown in
insets of Fig.2, the magnetization (M) increases linearly with H
applied along c-axis, then the M saturates above a critical magnetic
field ($H_c$) of about 1.5 T and 0.8 T for the crystals with x=0 and
0.15, respectively. Similar behavior is observed with H applied
within ab plane, the $H_c$ is about 1 T and 0.3 T for the crystals
with x=0 and 0.15, respectively. These results indicate that a
metamagnetic transition from antiferromagnetism (AFM) to
ferromagnetism (FM) happens with increasing magnetic field. It
should be pointed out that the M-H curve is linear before saturation
without loop for H$\parallel$c, while a step increase in M is
observed between linear behavior and saturation with a small loop
for H $\perp$c. The saturated magnetization for H $\perp$c is larger
than that for H$\parallel$c, eg: 7.01$\mu$$_B$ with H$\parallel$ab
and 6.71$\mu$$_B$ with H$\parallel$c for x=0 crystal. It indicates
that the easy axis is within ab plane. The high temperature
susceptibility data ($100K<T<300K$) were fitted by the Curie-Weiss
formulum: $\chi(T)=\chi_0+\frac{C}{T+\theta}$. $\chi$$_0$ is the
temperature-independent susceptibility, C is the Curie-Weiss
constant and $\theta$ is the Weiss temperature. The effective
magnetic moment and Weiss temperature are listed in Table I.
Effective magnetic moment is close to theoretical value of Eu$^{2+}$
ion: 7.94$\mu$$_B$. The Weiss temperature is negative, indicating a
ferromagnetic interaction between Eu$^{2+}$ ions. However, an
antiferromagnetic ordering occurs at low magnetic fields. A possible
magnetic structure is that the coupling of Eu$^{2+}$ ions is
ferromagnetic within ab plane, while antiferromagnetic for
interplane; that is: A-type antiferromagnetism. It needs neutron
experiment to confirm this speculation.

\begin{center}
\begin{table}[tp]%
\caption{\label{tab:table1}Magnetic parameters obtained by fitting
the high temperature (100 K$\sim$300 K) susceptibility data for
$Eu_{1-x}La_xFe_2As_2$ crystals with x=0 and 0.15 with Curie-Weiss
law: $\chi(T)=\chi_0+\frac{C}{T+\theta}$.}
\label{aggiungi} \centering %
\begin{tabular}{clccc}
\toprule %
EuFe$_2$As$_2$&H$\parallel$ab&H$\parallel$c\\\hline
$\theta$(K) & -24.67 & -21.60\\
C(emu/K Oe mol) & 7.78 & 7.52\\
$\mu$$_{eff}$($\mu$$_B$) & 7.89 & 7.76\\
\hline \hline
Eu$_{0.85}$La$_{0.15}$Fe$_2$As$_2$&H$\parallel$ab&H$\parallel$c\\
\hline $\theta$(K) & -22.82 & -22.20\\
C(emu/K Oe mol) & 7.86 & 7.45\\
$\mu$$_{eff}$($\mu$$_B$) & 7.93 & 7.72\\\hline\hline

\end{tabular}
\end{table}
\end{center}

\begin{figure}[t]
\includegraphics[width=9cm]{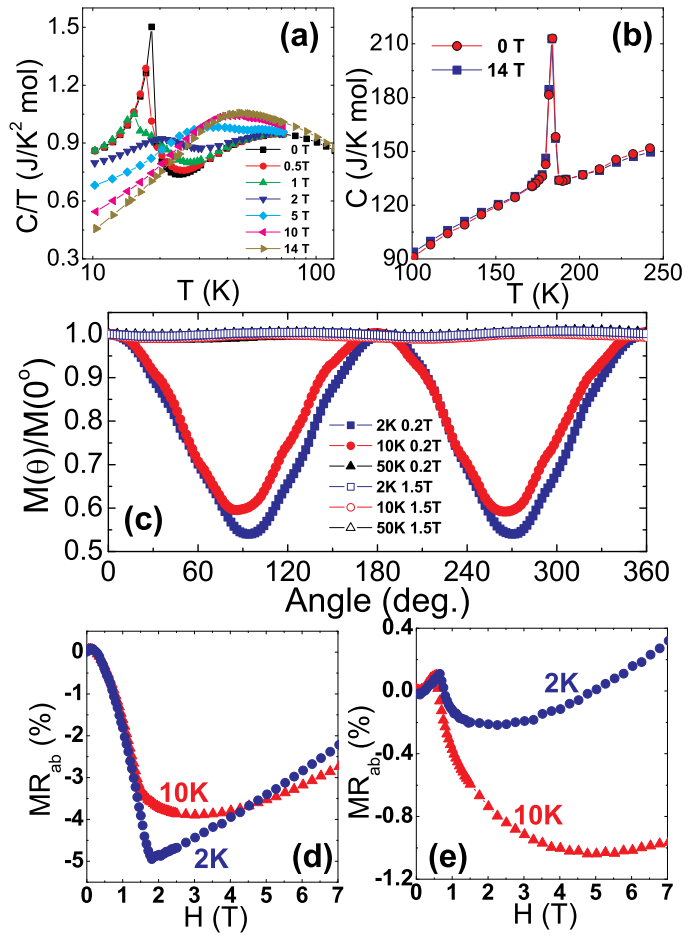}
\caption{(color online). Specific heat as a function of temperature
under H applied along c-axis for x=0 crystal, (a): below 100 K; (b):
above 100 K. (c). Angle dependent in-plane magnetization at 2, 10
and 50 K under H=0.2 T and 1.5 T, respectively; (d) and (e):
Isothermal in-plane magnetoresistance with H along c-axis for x=0
and x=0.15 crystals, respectively.
\\}
\end{figure}

In order to further study effect of H on magnetic ordering, specific
heat was measured with H applied along c-axis for x=0 crystal as
shown in Fig.3. A sharp jump around 185 K is observed, such anomaly
should arise from SDW/structural transition observed in resistivity.
Fig.3(b) shows no change for the anomaly at 185 K under H=14 T
relative to the case of H=0 T. It suggests that the effect of H=14 T
on the SDW/structural transition is negligible. Another jump around
20 K, associated with the magnetic ordering of Eu$^{2+}$ ions
observed in $\chi(T)$, shows up as shown in Fig.3(a). In contrast to
the anomaly related with SDW/structural transition at 185 K, the
jump associated with the magnetic ordering of Eu$^{2+}$ ions is
suppressed and shifts to low temperature with increasing H up to
about 1 T. When H$>$1T, the sharp jump becomes a broad peak and
shifts to high temperature with further increasing H. These results
are consistent with susceptibility behavior shown in Fig.2, and
further confirm that a metamagnetism from antiferromagnetism to
ferromagnetism occurs with increasing H. Similar behavior in
specific heat is observed in Na$_{0.85}$CoO$_2$ due to a
metamagnetic transtion\cite{luo}. In order to further understand
metamagnetism of Eu$^{2+}$ ions, the angular dependent magnetization
with rotating H within ab plane is measured for x=0 crystal. As
shown in Fig.3(c), an apparent twofold symmetry is observed at 10 K
and 2 K under H =0.2 T. The anisotropy is about 2.0 at 2 K. With
increasing \emph{T} to 50 K, the magnetization under H =0.2 T is
changed to be isotropic. It is intriguing that the magnetization is
also isotropic at 2 and 10 K under H=1.5 T. It suggests that the
magnetization in antiferromagnetic state is anisotropic in ab plane,
while in ferromagnetic state is isotropic. An interesting question
is naturally proposed: what makes them different? Magnetic structure
of BaFe$_2$As$_2$ is stripe-like AFM in Fe-As layer from neutron
scattering\cite{Huang}, a twofold magnetic symmetry at 4 K with
anisotropy of 1.14 has been reported\cite{Wang}. Therefore, it is
easily understood that the twofold symmetry is observed in
antiferromagnetic state of Eu$^{2+}$ ions. In ferromagnetic state,
the ferromagnetic arrangement of spins for $Eu^{2+}$ leads to a
large internal magnetic field. Such large internal magnetic field
can polarize the spin orientation in Fe-As layer, so that the
magnetization is isotropic in ferromagnetic state.

Fig.3(d) and (e) show the isothermal in-plane magnetoresistance (MR)
with H along c-axis at 2 and 10 K for x=0 and x=0.15 crystals,
respectively. Fig.3(d) shows that negative in-plane MR increases
with increasing H up to a certain magnetic field, then decreases
with further increasing H. The clear kink in in-plane MR at 2 K is
observed at $H\sim1.7$ T for x=0 crystal. As shown in Fig.3(e), the
in-plane MR is positive for x=0.15 crystal, and increases with
increasing H up to $\sim 0.7$ T, then decreases with further
increasing H. The magnetic field corresponding to the kink at 2 K is
almost the same as the critical magnetic field induced metamagnetic
transition observed in Fig.2. As shown in Fig.3(d) and (e), the kink
shifts to low magnetic field with increasing temperature. This is
easily understood because the kink is closely associated with
metamagnetic transition from AFM to FM. These results suggest that
there exist strong coupling between local moment of $Eu^{2+}$ and
charge in Fe-As layer.

\begin{figure}[t]
\includegraphics[width=9cm]{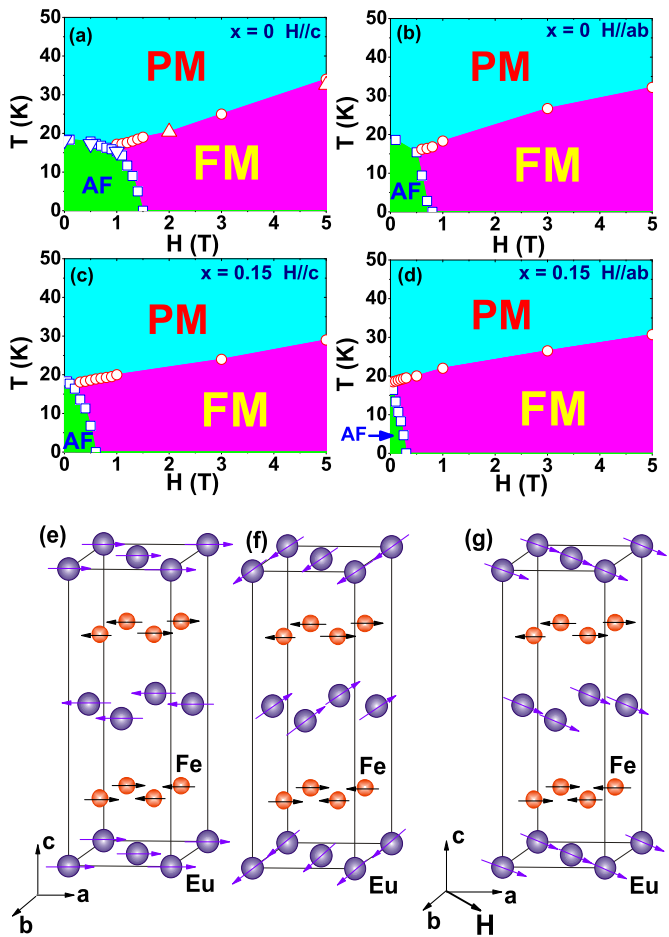}
\caption{(color online). Detailed H-T phase diagram for magnetism of
Eu$^{2+}$, (a): H$\parallel$c and (b): H$\parallel$ab for x=0
crystal; (c) H$\parallel$c and (d): H$\parallel$ab for x=0.15
crystal. The open squares stand for antiferromagnetic transition
temperature ($T_N$) determined by susceptibility. The open circles
present ferromagnetic transition temperature ($T_c$) determine by
d$\chi(T)$/dT. The open up-triangles and down triangles in (a)
present $T_c$ and $T_N$ determined by the peak position of specific
heat, respectively; (e) and (f): two possible AF magnetic
structures;  (f) Ferromagnetic structure induced by H applied along
(110) direction.
\\}
\end{figure}

Fig.4(a)-(d) show detailed H-T phase diagram for magnetism of
Eu$^{2+}$ ions for x=0 and x=0.15 crystals for H$\parallel$ab plane
and H$\perp$ab plane, respectively. The antiferromagnetic transition
temperature is determined by the kink in $\chi(T)$. The
ferromagnetic temperature is determined by the extremum in
$\frac{d\chi(T)}{dT}$. At low field, the magnetic structure is
A-type antiferromagnetism, while ferromagnetic state above critical
magnetic field. As shown in Fig.4(a)-(d), the critical field with H
along c-axis is two times of that with H applied within ab-plane for
both of the x=0 and 0.15 crystals. La doping suppresses the AFM
phase and leads to a decrease in the critical field. The critical
field with H along c-axis is about 1.5 T for x=0 and 0.8 T for
x=0.15 crystal. Possible magnetic structures for the spins of
$Eu^{2+}$ are proposed as shown in Fig.4(e) and 4(f) based on the
results of susceptibility and specific heat. In the possible
magnetic structures, the antiferromagnetic SDW in Fe-As layer keep
the same as that in BaFe$_2$As$_2$ determined by neutron
scattering\cite{Huang} since the different ions in Ba site have no
effect on magnetic structure of Fe$^{2+}$ for $MFe_2As_2$ (M=Ba, Sr,
and Ca). At low fields, the inter-plane coupling among Eu$^{2+}$
ions is antiferromagnetic, and the intra-plane coupling is
ferromagnetic; that is: A-type AFM structure for $Eu^{2+}$ spins.
The spin orientation of Eu$^{2+}$ ions has two possibilities
relative to the spin direction of $Fe^{2+}$. One possibility is that
the spin direction of Eu$^{2+}$ ions is perpendicular to that of
Fe$^{2+}$ ions, that is: noncollinear AFM structure, similar to that
of $Nd_2CuO_4$\cite{NCO1,NCO2,NCO3}. Another possibility is that the
spin direction of Eu$^{2+}$ ions is parallel to that of Fe$^{2+}$
ions, that is: collinear AFM structure. With increasing H, the
interplane coupling changes to ferromagnetic as shown in Fig.4(g).
Such interplane ferromagnetic coupling between $Eu^{2+}$ spins
enhances the coupling between Eu$^{2+}$ ions and Fe-As layer, so
that the internal magnetic field produced by FM of $Eu^{2+}$ has
strong effect on the SDW and the anisotropy disappears. The
understanding on interaction between magnetic ordering of rare earth
ions and SDW state of Fe-As layer is helpful to study the underlying
physics of Fe-As compound.

In summary, we systematically study the magnetic ordering of
$Eu^{2+}$ through the resistivity, susceptibility and specific heat
measurements in high-quality single crystal $Eu_{1-x}La_xFe_2As_2$.
A metamagnetic transition from antiferromagnetism to ferromagnetism
is found for magnetic sublattice of Eu$^{2+}$ ions. Detailed H-T
phase diagrams for x=0 and 0.15 crystals are given, and possible
magnetic structure for $Eu^{2+}$ spins is proposed. At low fields,
the magnetic structure of $Eu^{2+}$ spins is A-type
antiferromagnetic. There exists a strong coupling between local
moment of $Eu^{2+}$ and charge in conducting Fe-As layer. Our
results indicate a coupling between magnetism of rare earth ions and
SDW ordering in Fe-As layer. These intriguing phenomena from
magnetism of rare earth ions maybe shed light on the underlying
physics of FeAs superconductors.

\vspace*{2mm} {\bf Acknowledgment:} This work is supported by the
Nature Science Foundation of China and by the Ministry of Science
and Technology of China (973 project No: 2006CB601001) and by
National Basic Research Program of China (2006CB922005).

{\bf Note:} At completion of this work we became aware of one paper
reported susceptibility of $EuFe_2As_2$ by Jiang et al.,
arXiv:0808.0325.

\end{document}